\documentclass[twocolumn,showpacs,preprintnumbers,amsmath,amssymb]{revtex4}
\usepackage{graphicx}% Include figure files
\usepackage{bm}% bold math

\begin{document}

\title{Stabilization of impurity states in crossed
magnetic and electric fields}

\author{K. Krajewska and J. Z. Kami\'nski}
\affiliation{Institute of Theoretical Physics, Warsaw University,
Ho\.za 69, 00--681 Warszawa, Poland} 

\begin{abstract}
It is shown that the renormalizability of the zero-range interaction
in the two-dimensional space is always followed by the existence of a
bound state, which is not true for odd-dimensional spaces. A
renormalization procedure is defined and the exact retarded Green's
function for electrons moving in two dimensions and interacting with both
crossed magnetic and electric fields and an attractive zero-range
interaction is constructed. Imaginary parts of poles of this
Green's function determine lifetimes of quasi-bound (resonance)
states. It is shown that for some particular parameters the
stabilization against decay occurs even for strong electric fields.
\end{abstract}


%\date{\today}
\maketitle

%\newpage

\section{Introduction}
The observations of the integer and fractional quantum Hall effects
\cite{KDP80,TSG82} are among the most
important discoveries of recent years and have had a profound impact on
both applied and fundamental physics. Many aspects of these
discoveries are presented in the books edited by Prange  and Girvin
\cite{PG87} and by Das Sarma and Pinczuk \cite{SP97}, and in the
monographs \cite{CP95,E00}. From the theoretical point of view the
quantum Hall phenomenon is due to the two-dimensional dynamics of
electrons moving under the influence of crossed magnetic and electric
fields in the presence of impurities. Since the de Broglie wavelength
of electrons scattered by impurities is much larger then the
interaction range, therefore, it is legitimated
to describe a scatterer by a zero-range potential. This, in fact,
has been already proposed by Prange \cite{P81} shortly after the
discovery of the quantum Hall effect, and discussed further in
Refs. \cite{PJ82,JP84}. In his seminal paper Prange studied electron states in two
dimensions in perpendicular magnetic and in-plane electric fields in
the  presence of a single repulsive 'delta-like' scatterer. Prange's
model, with an attractive zero-range potential and zero electric
field, has been reconsidered by Perez and Coutinho \cite{PC91}, and by
Cavalcanti and de Carvalho \cite{CC98}. They showed that the
attractive zero-range interaction can be  rigorously defined by
renormalizing the strength of the $\delta$-function. Such a procedure
cannot be  develop for a repulsive two-dimensional $\delta$-function
(as it will follow shortly), for which an artificial cutoff has to be
introduced, as described for instance in Ref. \cite{HL99}. The problem of
electron scattering by a single  impurity represented by a repulsive
short-range potential in the presence of both magnetic and electric
fields has been further studied from the classical point of view in
Ref. \cite{BHHP96}, whereas the quantum  description has been developed in
Ref. \cite{HL99} for a repulsive, and in Ref. \cite{GM99} for an attractive
$\delta$-function potential. Both these quantum analyses show that due
to external magnetic and electric fields new long living quasi-bound
states appear in the positive part of the energy spectrum, the
existence of which appears to be crucial for  the explanation of the
robustness of the quantum Hall effect \cite{P81,HL99}.

There has been shown \cite{HL99,GM99} that in the limit of a very
weak electric field the lifetime of these quasi-bound states tends to
infinity and grow in a Gaussian way as the electric field tends to
zero. This finding, not taking into account, however, the precisely
determined functional dependence of the lifetime
on the strength of the electric field, agrees with our  common
understanding of the decay problem in a weak electric field. The aim
of this paper is to  analyze this problem for strong electric
fields. It is usually believed that with an increasing electric field
the lifetimes of these quasi-bound states should decrease to zero. It
appears, however, that such a behavior is not generally true and that
for some particular values of magnetic and  electric fields one
observes a new phenomenon, the stabilization of these new quasi-bound
states. We demonstrate this phenomenon for an attractive
$\delta$-function potential considered in Ref. \cite{GM99}.

In this paper we use units in which $\hbar=1$.

\section{Attractive $\delta$-function potential}
In this section we shall define the renormalized attractive
$\delta$-function potential for one-, two- and three-dimensional
cases. To this end let us assume that a quantum system without this
interaction is described by  the hamiltonian $H_0$. We introduce
further a regularization of a zero-range potential such that the
total hamiltonian is equal to
\begin{equation}
H=H_0+\lambda_{\sigma} \delta_{\sigma}
(\bm{r}),
\label{e2.1}
\end{equation}
in which $\sigma$ is a regularization parameter, $\lambda_{\sigma}$ is
a bare coupling constant, and  $\delta_{\sigma}(\bm{r})$ tends to
the Dirac distribution $\delta (\bm{r})$ in a $d$-dimensional
space. Our aim is to construct the retarded Green's function
satisfying the equation
\begin{equation}
(E-H)G^{(+)}(\bm{r},\bm{r}';E)=\delta(\bm{r}-\bm{r}'),
\label{e2.2}
\end{equation}
provided that the range of a regularized potential is
{\it very} small, i.e., is much smaller than the electron's de Broglie
wavelength or the size of a bound state wave function. This means
that in the  Lippmann-Schwinger equation
%\begin{widetext}
\begin{eqnarray}
& G^{(+)}(\bm{r},\bm{r}';E)= G_0^{(+)}(\bm{r},\bm{r}';E) \nonumber \\
& +\lambda_{\sigma}\int \text{d}^d\bm{r}'' 
G_0^{(+)}(\bm{r},\bm{r}'';E)\delta_{\sigma}(\bm{r}'')
G^{(+)}(\bm{r}'',\bm{r}';E),
\label{e2.3}
\end{eqnarray}
%\end{widetext}
in which $G_0^{(+)}(\bm{r},\bm{r}';E)$ is the Green's function for
the hamiltonian $H_0$, we can  approximate under the integral the full
Green's function $G^{(+)}(\bm{r}'',\bm{r}';E)$ by
$G^{(+)}(\bm{0},\bm{r}';E)$. This allows to calculate
$G^{(+)}(\bm{0},\bm{r}';E)$ and to arrive at the following
expression for the retarded Green's function,
%\begin{widetext}
\begin{eqnarray}
& G^{(+)}(\bm{r},\bm{r}';E)= G_0^{(+)}(\bm{r},\bm{r}';E) \nonumber \\
& +\frac{G_0^{(+)}(\bm{r},\bm{0};E) G_0^{(+)}(\bm{0},\bm{r}';E)} 
{\lambda^{-1}_{\sigma}-\int \text{d}^d\bm{r}'' 
G_0^{(+)}(\bm{0},\bm{r}'';E) 
\delta_{\sigma}(\bm{r}'')}.
\label{e2.4}
\end{eqnarray}
%\end{widetext}
In order to proceed further let us choose a particular regularization
prescription for which all the space  and momentum integrals, that
will appear below, can be carried out and expressed in terms of
elementary functions, namely,
\begin{equation}
\delta_{\sigma}(\bm{r})=\int \frac{\text{d}^d\bm{k}}{(2\pi)^d}
\exp(\text{i}\bm{k}\cdot\bm{r} -\sigma^2\bm{k}^2).
\label{e2.5}
\end{equation}
Let us also assume for a moment that $H_0$ describes free electrons of
a reduced mass $m^*$. For such a hamiltonian the retarded Green's
function adopts the form
\begin{widetext}
\begin{equation}
G_0^{(+)}(\bm{r},\bm{r}';E)=-\text{i}\int_0^{\infty}
\text{d}t \text{e}^{\text{i}(E+\text{i}\varepsilon)t}
\int \frac{\text{d}^d\bm{k}}{(2\pi)^d} \exp\biggl (
\text{i}\bm{k}\cdot (\bm{r} -\bm{r}')
-\text{i}\frac{\bm{k}^2}{2m^*}t \biggr ),
\label{e2.6}
\end{equation}
\end{widetext}
in which $\varepsilon$ is an infinitesimally small positive real
number. Thus, after performing the  Gauss integration over $\bm{k}$,
the denominator in eq. (\ref{e2.4}) becomes
\begin{equation}
D(E)= \lambda^{-1}_{\sigma}+\text{i}\biggl (
\frac{m^*}{2\pi\text{i}}\biggr )^{d/2}
\int_0^{\infty} \text{d}t \frac{\text{e}^{\text{i}(E+\text{i}\varepsilon)t}}
{(t-2\text{i}m^*\sigma^2)^{d/2}}.
\label{e2.7}
\end{equation}
As we see, for $d=1$ the last integral exists for $\sigma=0$, i.e.,
there is no need for the renormalization of a coupling
constant $\lambda_{\sigma}$. On the other hand, for $d>1$ the integral
in the equation above diverges for small $t$ in the limit of a vanishing
regularization parameter $\sigma$. It can be shown, however, that the
divergent part  of this integral can be absorbed into the bare
coupling constant $\lambda_{\sigma}$. To this end let us consider in
the beginning the case of $d=2$. Integrating by parts we end up  in
the limit of vanishing $\sigma$ with
\begin{equation}
D(E)=\lambda_R^{-1}+\frac{m^*}{2\pi} \biggl ( \ln(\text{i})
-\text{i}E\int_0^{\infty} \text{d}t 
\text{e}^{\text{i}(E+\text{i}\varepsilon)t}\ln(t/t_0) \biggr ),
\label{e2.8}
\end{equation}
where $t_0$ is an arbitrary positive real number of the same
dimensionality as $t$, introduced only for dealing with a
dimensionless argument of the logarithm function, and $\lambda_R$  is
the renormalized coupling constant
\begin{equation}
\lambda_R^{-1}=\lim_{\sigma\rightarrow 0} \Bigl [ 
\lambda_{\sigma}^{-1}-\frac{m^*}{2\pi}\ln\Bigl (
\frac{2m^*\sigma^2}{t_0}\Bigr )  \Bigr ].
\label{e2.9}
\end{equation}
As it follows from the equation above the bare coupling constant
$\lambda_{\sigma}$ has to be negative in order to carry out the
renormalization procedure. Hence, the zero-range limit exists only for
an attractive interaction. Performing the remaining integration we
finally arrive at
\begin{equation}
D(E)=\lambda_R^{-1}+\frac{m^*}{2\pi}\ln\biggl ( 
\frac{\text{e}^{-\gamma}}{-Et_0} \biggr ),
\label{e2.10}
\end{equation}
where $\gamma$ is the Euler's constant. Since the energy of a bound
state, $E_B<0$, supported by  this interaction is determined by the zero
of $D(E)$, therefore
\begin{equation}
E_B=-\frac{1}{t_0}\text{e}^{-\gamma+2\pi/\lambda_Rm^*}.
\label{e2.11}
\end{equation}
As one sees, the expression above is always negative, independently
weather the renormalized  coupling constant is negative or
positive. This means that the zero-range potential in two-dimensional
space always supports one bound state. Combining (\ref{e2.10}) with
(\ref{e2.11}) we arrive at the following expression for the
denominator $D(E)$,
\begin{equation}
D(E)=\frac{m^*}{2\pi}\ln\biggl (\frac{E_B}{E} \biggr ),
\label{e2.11a}
\end{equation}
in which an artificial parameter $t_0$ does not appear any more.

\begin{figure*}
\includegraphics[width=11cm]{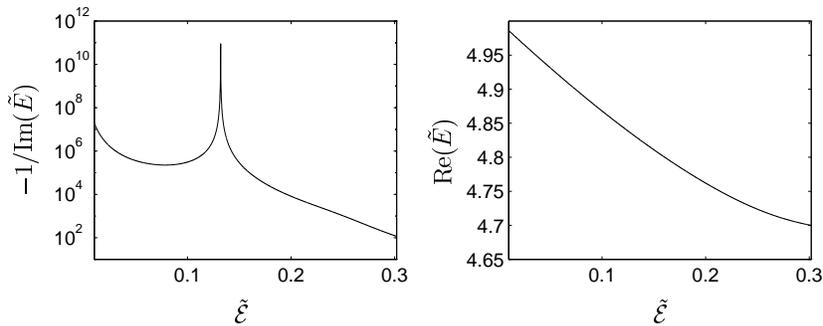}
\caption{\label{f1}
Presents the dependence of the lifetime $\tau=-1/\text{Im}(\tilde{E})$ and
the real part of $\tilde{E}$ on the scaled electric field $\tilde{\mathcal{E}}$
for an impurity which supports one bound state of energy $\tilde{E}_B=-6.4$.
For $\tilde{\mathcal{E}}=0.132$ we observe the stabilization of a resonance state
located just below the third Landau level.
}
\end{figure*}

\begin{figure*}
\includegraphics[width=11cm]{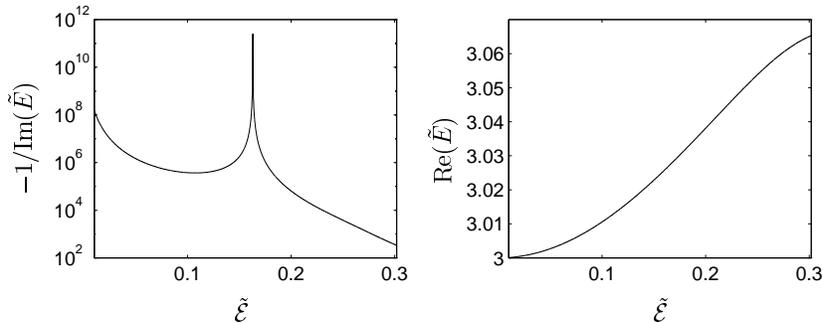}
\caption{\label{f2}
The same as in Fig. \ref{f1} but for $\tilde{E}_B=-2.8$. The stabilization
occurs for $\tilde{\mathcal{E}}=0.163$ for a resonance state located just above the second Landau level.
}
\end{figure*}

\begin{figure*}
\includegraphics[width=11cm]{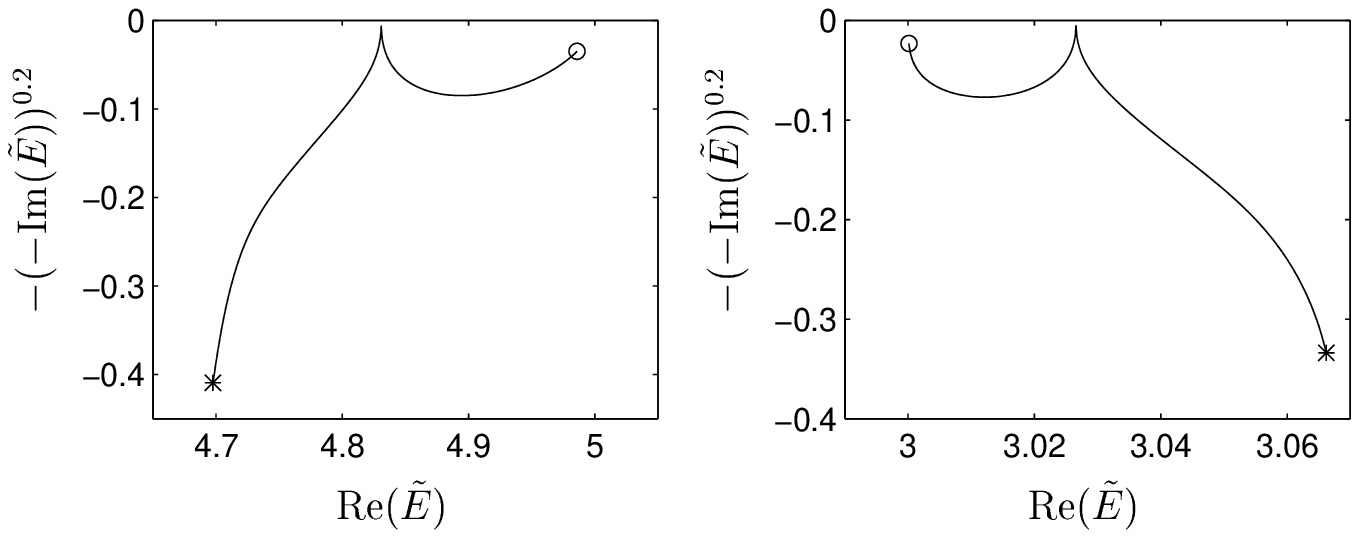}
\caption{\label{f3}
Shows the trajectories of two resonances considered in Figs. \ref{f1} and
\ref{f2} in the complex energy plane. For the visual purpose the imaginary
part is raised to the power $1/5$. The left frame corresponds to the binding
energy $\tilde{E}_B=-6.4$ and the right one to $\tilde{E}_B=-2.8$. The circle
marks the minimum value $\tilde{\mathcal{E}}=0.01$, whereas the asterisk indicates
the maximum value $\tilde{\mathcal{E}}=3.1$. The stabilization occurs for
$\tilde{\mathcal{E}}=0.132$, $\text{Re}(\tilde{E})=4.83$,
$\text{Im}(\tilde{E})=-0.111\times 10^{-10}$ and
$\tilde{\mathcal{E}}=0.163$, $\text{Re}(\tilde{E})=3.0266$,
$\text{Im}(\tilde{E})=-0.398\times 10^{-11}$, respectively.
}
\end{figure*}

The existence of a bound state for a zero-range interaction,
irrespectively of the sign of a renormalized coupling constant, can
only happen for $d=2$. In  order to show this let us consider the case
of $d=3$. Now we do not need to introduce $t_0$ and end up with
\begin{equation}
D(E)=\lambda_R^{-1}+\frac{m^*}{2\pi}\sqrt{-2m^*E},
\label{e2.12}
\end{equation}
where the renormalized coupling constant is equal to
\begin{equation}
\lambda_R^{-1}=\lim_{\sigma\rightarrow 0} \Bigl ( 
\lambda_{\sigma}^{-1}+\frac{m^*}{2\pi^{3/2}\sigma} 
\Bigr ) .
\label{e2.13}
\end{equation}
Hence, it is clearly seen that again the renormalization procedure can
be only performed for a negative  bare coupling
constant. Nevertheless, contrary to the two-dimensional case, a bound
state exists only for a negative renormalized coupling constant
$\lambda_R$. A similar situation occurs for the one-dimensional
$\delta$-function potential supporting one bound state
just for negative coupling constants (renormalized and bare coupling
constants are equal here and amount to $-\sqrt{2|E_B|/m^*}$). 

Summarizing, for the two-dimensional zero-range interaction the
renormalizability is followed
by the existence of a bound state, 
irrespectively of the sign of the renormalized coupling constant. In our 
further discussion we shall limit ourselves to the two-dimensional
renormalizable zero-range interaction. Let us also note in closing this 
section that the zero-range potential for an arbitrary dimensional space 
has been studied by W\'odkiewicz \cite{W90} who also emphasized the 
difference between odd and even dimensions.

\section{Quasi-bound states in crossed magnetic and electric fields}
We can now include into our two-dimensional model external magnetic and
electric fields. For this we need the exact form of the retarded Green's
function for the hamiltonian
\begin{equation}
H_0=\frac{1}{2m^*}\Bigl ( -\text{i}\bm{\nabla}-e\bm{A}(\bm{r})
\Bigr )^2-e\bm{\mathcal{E}}\cdot\bm{r} .
\label{e3.1}
\end{equation}
Let us assume that electrons move in the $xy$-plane, the magnetic field
$\bm{\mathcal{B}}$ is perpendicular to this plane and the electric field 
$\bm{\mathcal{E}}$ points into the $x$-direction. For such a geometry and 
for the vector potential in the symmetric gauge, i.e. 
$\bm{A}=\frac{1}{2}\bm{\mathcal{B}}\times\bm{r}$,
\begin{equation}
H_0=\frac{1}{2m^*}\Bigl[ \Bigl( -\text{i}\partial_x+\frac{e{\mathcal{B}}}{2}y\Bigr )^2+
\Bigl(-\text{i}\partial_y-\frac{e{\mathcal{B}}}{2}x\Bigr )^2\Bigr ]-e{\mathcal{E}}x,
\label{e3.2}
\end{equation}
and the Green's function adopts the form
\begin{widetext}
\begin{eqnarray}
G_0^{(+)}(\bm{r},\bm{r}';E)=
-\frac{m^*\omega}{4\pi}\int_0^{\infty} 
\text{d}t \frac{\text{e}^{\text{i}(E+\text{i}\varepsilon)t}}
{\sin\frac{\omega t}{2}}
\exp\biggl[ \frac{\text{i}m^*\omega}{4}\bigl( (x-x')^2
+(y-y')^2\bigr )\cot\frac{\omega t}{2}\nonumber\\
+\frac{\text{i}m^*\omega}{2}(xy'
-x'y)+\frac{\text{i}e\mathcal{E}t}{2}(x+x')
+\frac{\text{i}e{\mathcal{E}}}{\omega}\Bigl( \frac{\omega t}{2}
\cot\frac{\omega t}{2}-1\Bigr )
\Bigl( -y+y'+\frac{e{\mathcal{E}}t}{2m^*\omega}\Bigr )\biggr ],
\label{e3.3}
\end{eqnarray}
\end{widetext}
with $\omega=|e|{\mathcal{B}}/m^*$ being the cyclotron frequency. Since
lifetimes of quasi-bound states are determined by poles of the Green's
function, therefore, we shall concentrate in our further discussion 
on the denominator in eq. (\ref{e2.4}). One can find that in this 
particular case $D(E)$ can be expressed as follows,
\begin{widetext}
\begin{eqnarray}
D(E)=\lambda_{\sigma}^{-1}+\frac{m^*\omega}{4\pi}\int_0^{\infty} 
\text{d}t \frac{\text{e}^{\text{i}(E+\text{i}\varepsilon)t}}
{\sin\frac{\omega t}{2}-\text{i}m^*\omega{\sigma}^2\cos\frac{\omega t}{2}}
\exp\Biggl[ \frac{\text{i}t}{2m^*}\Bigl( \frac{e{\mathcal{E}}}{\omega}\Bigr )^2
\Bigl( \frac{\omega t}{2}\cot\frac{\omega t}{2}-1\Bigr )\nonumber\\
-{\sigma}^2\Bigl( \frac{e{\mathcal{E}}}{\omega}\Bigr )^2
\frac{\bigl( \frac{\omega t}{2}\bigr )^2+
\bigl( \frac{\omega t}{2}\cot\frac{\omega t}{2}-1\bigr )^2}
{1-\text{i}m^*\omega{\sigma}^2\cot\frac{\omega t}{2}}\Biggl ],
\label{e3.4}
\end{eqnarray}
\end{widetext}
where the integral above diverges logarithmically for
small values of $t$ in the zero-range limit. However, as we have proved
it in the previous section, this singularity can be removed by combining 
the divergent term  with the bare coupling constant $\lambda_\sigma$.

For our further analysis it is convenient to introduce dimensionless
variables, in which the length is measure in the units of
$1/\sqrt{m^*\omega}$, whereas the electric field and the energy
are scaled as  $\tilde{\mathcal{E}}=|e|{\mathcal{E}}/\sqrt{m^*\omega^3}$ and
$\tilde{E}=2E/\omega$, respectively. Changing the integration
variable to $s=\omega t/2$ we obtain finally for $\sigma\rightarrow 0$,
\begin{widetext}
\begin{equation}
 D(E)=\frac{m^*}{2\pi}\biggl [
\ln\biggl( \frac{\tilde{E}_B}{\tilde{E}}\biggr ) 
 +\int_0^{\infty}\text{d}s \text{e}^{\text{i}(\tilde{E}+
\text{i}\varepsilon)s}\Biggl( \frac{\exp\bigl( \text{i}
{\tilde{\mathcal{E}}}^2 s(s\cot s-1)\bigr )}{\sin s}-\frac{1}{s}\Biggr )
\biggr ].
\label{e3.5}
\end{equation}
\end{widetext}

We shall demonstrate now that even in strong electric fields
one can observe long living quasi-bound states. This nonperturbative
result follows from the numerical determination of zeros of $D(E)$. It
appears that for some particular values of the electric field and the
binding energy there exist zeros of (\ref{e3.5}) with very small
imaginary parts, although real parts remain positive. This phenomenon,
which we call the stabilization, is presented in Figs.  \ref{f1},
\ref{f2} and \ref{f3}. In Figs. \ref{f1} and \ref{f2} we draw the
dependence of lifetimes,
$\tau=-(2\text{Im}(E))^{-1}=-(\omega\text{Im}(\tilde{E}))^{-1}$, and
real parts of $E$, $\text{Re}(E)=\frac{\omega}{2}\text{Re}(\tilde{E})$, in
units of $\omega^{-1}$ and $\omega/2$, respectively. We see that with
the increasing electric field the lifetime initially decreases, which
is a commonly accepted result. We observe, however, that at a
particular value for the electric field the lifetime starts
increasing, approaches its maximum, and then monotonically decreases.
In both presented cases the real part of $E$ remains positive and is just
below the third Landau level of energy $5\omega/2$ or above the second
one of energy $3\omega/2$.  In Fig. \ref{f3} we show the position of
these poles of the Green's function in the complex energy plane. Only
for the visual purpose the imaginary part is raised to the power
$1/5$. We see that with an increasing electric field the poles in the
beginning depart from the real axis, but afterward start approaching
it, reach minimum for the absolute value of the imaginary part of $E$ (which
appears to be almost equal to zero within the accuracy of our
numerical calculation) and then again migrate downward. Such an
unexpected non-monotonic behavior happens for some particular values
of the binding energies as well as the applied fields. At this point let us
only mention that estimations of magnetic and electric fields show that 
they can be generated easily in experimental setups. 

What we have checked in our numerical investigations is that the 
stabilization occurs only for states which are close to the excited Landau 
levels and does not appear for states near the first Landau level. Indeed, 
for very small electric fields resonances considered
in this paper approach the second and the third Landau levels, i.e.,
the excited ones. This appears to be a general rule that the electric field
generates new resonance states in a close vicinity of {\em excited}
Landau levels. These states differ from the ones considered by Prange
\cite{P81} and reinvestigated by Cavalcanti and de Carvalho \cite{CC98}.
In their case, when the electric field is switched off, the point 
interaction modifies only Landau states with the
vanishing angular momentum, because only for these states the wavefunction
does not vanish at the origin. The states with a non-zero angular momentum,
i.e., the vortex Landau states, for which the wavefunctions explicitly
depend on the polar angle $\varphi$ and vanish at the origin, are not affected
by a point interaction. The situation changes if a small electric field
is applied. Then the vortex Landau states acquire small contributions
which do not vanish at the origin, as follows for instance from perturbation 
theory. Hence, the interaction with a contact potential does not vanish, and 
apart from the well-known states considered previously \cite{P81,CC98} 
new resonance states emerge close to the excited Landau levels. These are 
the states for which the stabilization takes place. Since the wavefunction
of them for small electric fields is predominantly composed of the vortex 
Landau states of non-zero angular momentum, therefore, one can call them
the vortex resonance states. This fact suggests that the stabilization 
presented in this paper is due to the angular motion of electrons around 
the impurity, as it also happens in the classical considerations for a 
repulsive potential \cite{BHHP96}. The analysis presented above can be 
applied to weak electric fields. For strong nonperturbative electric fields
we can rather rely on the exact numerical analysis of this problem with the 
hope that the picture above still remains valid. However, the almost total 
suppression of ionization rates for some particular values of the electric 
field hardly can be explained only in terms of classical physics and we 
expect that its origin is due to complicated quantum interference effects 
which can only be studied numerically.

\begin{acknowledgments}
This work has been supported by the Polish Committee for Scientific Research
(Grant No. KBN~2~P03B 039~19).
\end{acknowledgments}

\end{document}